\RequirePackage[2020-02-02]{latexrelease}
\documentclass[reprint,
superscriptaddress,
 amsmath,amssymb,
 aps,prd,twocolumn
]{revtex4}

\usepackage{graphicx}
\usepackage{dcolumn}
\usepackage{bm}

\begin{document}
\title{Implications of quantum gravity for dark matter in the brane-world scenario}
\author{Cao H. Nam}
\email{nam.caohoang@phenikaa-uni.edu.vn}  
\affiliation{Phenikaa Institute for Advanced Study and Faculty of Fundamental Sciences, Phenikaa University, Yen Nghia, Ha Dong, Hanoi 12116, Vietnam}
\date{\today}

\begin{abstract} 
Based on the swampland program establishing the constraints that an effective field theory must satisfy in order to come from quantum gravity, we point to that if this program is true the brane-world scenario with the branon dark matter (DM) would be ruled out without needing the experimental observations. In other words, the constraints of quantum gravity imply that the branons must be absorbed by the Kaluza-Klein (KK) gauge bosons which are the off-diagonal components of the bulk metric to become their longitudinal modes. Interestingly, the KK gauge bosons behave as the DM and hence it leads to a geometric unification of gravity and the DM, as a natural feature of the brane-world scenario completed into quantum gravity in the ultraviolet. In addition, the KK gauge boson DM would open a particularly promising observation window to search for the DM coupling very weakly to the SM in the cosmic microwave background and the spectrum of the primordial gravitational waves.
\end{abstract}

\maketitle

\section{Introduction}
The brane-world scenario has been an attractive paradigm for the extension of general relativity (GR) and standard model (SM) in the range of high enough energy, which is motivated by superstring/M theory (regarded as a consistent theory of quantum gravity) and provides the interesting phenomenological models for cosmology, astrophysics, and particle physics. (See \cite{Maartens2010} for a review.) From the string theory description, the SM fields are the excitations of the open strings which end in general on a stack of the D$p$-branes wrapping around a compact manifold of $(p-3)$ dimensions. Whereas, only the close strings which describe the gravitational sector can freely propagate in the whole spacetime or the bulk. Accordingly, the brane-world models can be realized from the compactification solutions of superstring/M theory which lead to the effective low-energy theories with the extra dimensions and the presence of a $3$-brane on which the SM fields are confined \cite{Horava1996,Lukas1999}. 

An important phenomenological consequence of the brane-world scenario is that the branons describing the fluctuations of the $3$-brane along the extra dimensions are a candidate for the dark matter (DM) \cite{Cembranos2003} which constitutes approximately $85\%$ of the matter in the universe but whose features (i.e. mass, stability, production mechanism, and couplings to the SM particles) remain mysterious and are of the major questions in particle physics and cosmology. (It may also solve the hierarchy problem \cite{Arkani-Hamed1998,Randall1999,Nam2021} and provide the models for the late accelerating phase \cite{Wong2010}.) The branons are the Goldstone bosons associated with the spontaneous breaking of the isometry in the extra dimensions due to the presence of the $3$-brane. It was argued in \cite{Cembranos2003} that if the brane tension is much lower than the higher-dimensional Planck scale, the particle spectrum at the range of the low energy is the branons and the SM particles. In particular, because the branons are stable by the parity on the brane and weakly interacting, they would be an interesting and natural candidate for the DM. The branon DM has been a particular class of weakly interacting massive particles \cite{Baer2015} and its signals have been detected in various experiments like LHC \cite{Cembranos2004,Prado2011,Cembranos2013}, LEP \cite{L3-Collaboration}, gamma-ray telescopes \cite{Strigari2008,Gammaldi2012,Cembranos2012,Nieto2020}, white dwarf stars \cite{Panotopoulos2017}, and dwarf spheroidal galaxies \cite{Miener2022}.

The connection between quantum gravity with particle physics and cosmology has recently become one area of active and important research. Quantum gravity arguments, string theory compactification, and microscopic physics have established together the swampland program that determines the universal properties of quantum gravity leading to the constraints that an effective field theory (EFT), like the brane-world scenario or the SM, must satisfy to be consistent with quantum gravity in the ultraviolet (UV) \cite{Vafa2005,Ooguri2007,Palti2019}. Hence, the swampland program provides not only new guiding principles to solve the open problems in particle physics and cosmology \cite{Martin-Lozano2017,Valenzuela2017,Hamada2017,Gonzalo2018a,Gonzalo2018b,Nam2022a,Nam2022b} but also UV imprints of quantum gravity at low-energy scales which open a window into understanding quantum gravity.

In this work, we study the DM problem in the brane-world scenario within the swampland program. We show that the branon DM violates at least three essential swampland conjectures and as a result, the brane-world scenario with the branon DM cannot be consistently coupled to quantum gravity or cannot come from the string theory compactification. Thus, the swampland program if true would allow us to rule out the branon DM without needing the experimental measurements. On the other hand, the branons cannot appear in the low-energy particle spectrum, but they would be absorbed by the Kaluza-Klein (KK) gauge fields which are the off-diagonal components of the bulk metric. Interestingly, we show that these KK gauge fields play the role of the DM and they lead to new implications for the search for the DM in the cosmic microwave background (CMB) as well as the spectrum of the primordial gravitational waves (PGWs). Indeed, because of their geometric nature and gravitational origin, they would typically couple very weakly to the particles in the SM and thus they actually behave as the DM. In this way, quantum gravity implies a geometric unification of gravity and the DM.

\section{The brane-world model}
We work in a simple model which consists of the $3$-brane propagating in the 5D bulk $\mathcal{M}_5$. Note that, the following analyses are qualitatively similar for the case of the more dimensional bulk spacetime which is compactified on a torus $T^n$. The corresponding action is given as follows
\begin{eqnarray}
S&=&S_{\text{bulk}}+S_{\text{brane}},\label{totact}\\
S_{\text{bulk}}&=&\frac{M^3_5}{2}\int d^5X\sqrt{-G}\left(\mathcal{R}_5-2\Lambda_5\right),\label{bulkact}\\
S_{\text{brane}}&=&\int d^4x\sqrt{-\widetilde{g}}\left(-f^4+\mathcal{L}_{\text{SM}}+\cdots\right),\label{braneact}
\end{eqnarray}
where $M_5$ is the 5D Planck scale, $\mathcal{R}_5$ is the 5D scalar curvature, $\Lambda_5$ is the bulk cosmological constant that is negative as seen later, $X^M(x^\mu)$ are the coordinates defined on the bulk($3$-brane), $G$($\widetilde{g}$) is the determinant of the metric on the bulk($3$-brane), $f$ is to the brane tension parameter, $\mathcal{L}_{\text{SM}}$ denotes the Lagrangian of the SM, and the ellipsis denotes the fields beyond the SM like the inflaton.

Considering $\mathcal{M}_5$ compactified on a circle $S^1$ and using the KK decomposition, we can split the bulk metric into various 4D component fields as follows
\begin{eqnarray}
ds^2_{\text{bulk}}&=&G_{MN}dX^MdX^N,\nonumber\\
&=&g_{\mu\nu}dx^\mu dx^\nu-R^2\left(d\theta+X_\mu dx^\mu\right)^2,\label{KKmetric}
\end{eqnarray}
where $g_{\mu\nu}$, $X_\mu$, and $R$ are the 4D tensor, 4D vector, and scalar components of the bulk metric, respectively, and $X^M=\left(x^\mu,\theta\right)$ are the bulk coordinates with $\theta$ to be an angle parametrizing $S^1$. The position of the $3$-brane in the bulk is given in the static gauge as $Y^M(x)=(x^\mu,Y^\theta(x))$ where $Y^\theta(x)$ is the branon describing the fluctuations of the $3$-brane along $S^1$ \cite{Sundrum1999}. In the following, we assume that the radion field $R$ is stabilized as $\langle R\rangle\equiv R_0$. From the ansatz of the bulk metric, one can find the induced metric on the $3$-brane as $ds^2_{\text{brane}}=\widetilde{g}_{\mu\nu}dx^\mu dx^\nu$ where $\widetilde{g}_{\mu\nu}$ is determined by
\begin{eqnarray}
\widetilde{g}_{\mu\nu}&=&G_{MN}\partial_\mu Y^M\partial_\nu Y^N,\nonumber\\
&=&g_{\mu\nu}-R^2_0\left(X_\mu+\partial_\mu Y^\theta\right)\left(X_\nu+\partial_\nu Y^\theta\right).\label{indmetr}
\end{eqnarray}

The isometry or the translation invariance on the circle $S^1$ implies the following $U(1)$ local transformation
\begin{eqnarray}
\theta&\longrightarrow&\theta+\alpha(x),\nonumber\\
Y^\theta&\longrightarrow&Y^\theta+\alpha(x),\label{GauTr}\\
X_\mu&\longrightarrow&X_\mu-\partial_\mu\alpha(x).\nonumber
\end{eqnarray}
Note that, the transformation of $\theta$ and $Y^\theta$ is the same because $Y^\theta$ determines the position of the $3$-brane along $S^1$. The transformation (\ref{GauTr}) means that the 4D vector component of the bulk metric behaves as a gauge field which describes the excitation of the 5D spacetime from the ground state $\mathcal{M}_4\times S^1$. We can show that the bulk metric and the induced metric on the $3$-brane are invariant under the transformation (\ref{GauTr}) and thus the total action (\ref{totact}) is too.

The dimensional reduction of the bulk action (\ref{bulkact}) on $S^1$ leads to an effective action for 4D Einstein gravity and 4D massless Abelian gauge field as
\begin{eqnarray}
S_{4D}\supset\int d^4x\sqrt{g}\left(\frac{M^2_{\text{P}}}{2}\mathcal{R}_4-\frac{1}{4\kappa^2}X_{\mu\nu}X^{\mu\nu}\right),\label{4Dgravact}
\end{eqnarray}
where $g\equiv\text{det}(g_{\mu\nu})$, $\mathcal{R}_4$ is the 4D scalar curvature which is a function of the 4D metric $g_{\mu\nu}$, $M_{\text{P}}$ is the observed Planck scale identified as $M^2_{\text{P}}\equiv 2\pi R_0M^3_5$, and $\kappa$ is the KK gauge coupling given as 
\begin{eqnarray}
\kappa\equiv\sqrt{2}\frac{R^{-1}_0}{M_{\text{P}}}.\label{KKcoup}
\end{eqnarray}

\section{Branon versus KK gauge boson}

From the induced metric (\ref{indmetr}) on the 3-brane, we see that the branon playing the role of the DM found in \cite{Cembranos2003} can appear in the low-energy particle spectrum in the limit that the KK gauge field $X_\mu$ decouples from the theory or in other words the local isometry of $S^1$ as given by (\ref{GauTr}) becomes global. On the contrary, it would be eaten to become the longitudinal component of $X_\mu$ \cite{Dobado2001,Kang2001}. In this limit, the branon is the physical excitation whose mass is vanishing or nonzero if the global isometry is realized as an exact or approximate symmetry, respectively.

Now, we use the swampland conjectures to show that the decoupling limit of the KK gauge field is inconsistent with quantum gravity. First, we observe that the decoupling limit of the KK gauge field can be realized when its kinetic term becomes infinite which would stop its dynamical propagation and hence is derived by sending the KK gauge coupling $\kappa$ to zero as seen in Eq. (\ref{4Dgravact}). The vanishing limit of $\kappa$ suggests $R_0\rightarrow\infty$ or the decompactification limit as seen in Eq. (\ref{KKcoup}). However, the $R_0\rightarrow\infty$ limit would violate at least the following three swampland conjectures:
\begin{itemize}
    \item The Sublattice Weak Gravity Conjecture (WGC) \cite{Reece2017,Reece2018} imposes an upper bound on the cut-off $\Lambda_{\text{UV}}$ of the EFT as $\Lambda_{\text{UV}}\lesssim\kappa^{1/3}M_{\text{P}}$. This means that the cut-off $\Lambda_{\text{UV}}$ of the EFT would approach zero in the decoupling limit of the KK gauge field or the limit of $R_0\rightarrow\infty$  and consequently it invalidates the description of the EFT.
    \item The mass of the KK modes can be expressed as follows
\begin{eqnarray}
m_n=\frac{n}{R_0}\sim e^{-d(R_i,R_0)},
\end{eqnarray}
where $R_i$ is a finite radius and $d(R_i,R_0)$ is the distance between two points $R_i$ and $R_0$ in the field space of the radion. In the limit of $R_0\rightarrow\infty$ corresponding to $d(R_i,R_0)\rightarrow\infty$, it yields an infinite tower of exponentially light states, as stated by the Swampland Distance Conjecture (SDC) \cite{Ooguri2007} which signals the breakdown of the EFT because it is impossible to describe an EFT coupled weakly to gravity with infinitely many light states.
\item The cosmological and astrophysical observations have indicated a de Sitter (dS) universe. Hence, the mass of the KK modes should satisfy the Festina Lente (FL) bound \cite{Montero2020,Vafa2021} as $m^2_n\gtrsim\sqrt{6}\kappa nM_{\text{P}}H_0$ with $H_0$ to be the observed Hubble parameter today. It is clear that this bound would be violated when taking $R_0\rightarrow\infty$ corresponding to the decoupling limit of the KK gauge field.
\end{itemize}
Furthermore, the vanishing limit of the KK gauge coupling would lead to trouble with the black hole entropy bound \cite{Bekenstein1981}: the evaporation of large black holes with all possible charges until reaching the extremal limit would result in a large number of remnants $N\sim1/\kappa$ which the difference in the mass is $\Delta M\sim\kappa M_{\text{P}}$; since the limit of $\kappa\rightarrow0$ leads to $N\rightarrow\infty$ and $\Delta M\rightarrow0$; as a result, there are an infinite number of different states which correspond to the same black hole geometry configuration and hence the black hole entropy is infinite in contradiction with the expectation that the black hole entropy which is proportional to the square of its mass should be finite. These results are deeply related to the no-global symmetries conjecture in quantum gravity which states that there are no exact global symmetries in the theories coupled to quantum gravity with a finite number of states \cite{Banks2011}.

Indeed, the violation of the above swampland conjectures in the decoupling limit of the KK gauge field is completely compatible with the observational problems. This limit leads to the light KK graviton modes which would significantly change the usual inverse-square law of the gravitational force whose modified potential is conventionally parametrized as $V(r)/V_N(r)=(1+\alpha e^{-r/R_0})/r$ where $V_N(r)$ is the Newtonian gravitational potential and $\alpha=2$ is the number of the first KK graviton modes. The fifth-force constraints impose an upper bound on the radius of $S^1$ as $R_0\lesssim \mathcal{O}(40)$ $\mu$m \cite{Kapner2007,Murata2015}. Also, the light KK graviton states can appear as missing energy/momentum in the colliding processes \cite{Giudice1999,THan1999}. Additionally, they can lead to the cross-section enhancement of the scattering processes like $e^-e^+\rightarrow f^-f^+$ above the predictions of the SM which have been well tested by the experimental measurements. Other seriously considerable deviations due to the presence of the light KK graviton modes can come from the Supernova cooling \cite{Cheng2010}, the expansion of the universe during the era of Big-Bang Nucleosynthesis (BBN) \cite{Cheng2010}, and the black hole production at the colliders \cite{Dimopoulos2001}.

The above arguments imply that the presence of the KK gauge boson associated with the local isometry of $S^1$ is avoidable to make the EFT to be consistently coupled to quantum gravity and compatible with the experimental observations. Using the expansion $\sqrt{-\widetilde{g}}=\sqrt{-g}(1-R^2_0|X_\mu+\partial_\mu Y^\theta|^2/2+\cdots)$ where the ellipse refers to the higher-order terms of $|X_\mu+\partial_\mu Y^\theta|^2\equiv g^{\mu\nu}(X_\mu+\partial_\mu Y^\theta)(X_\nu+\partial_\nu Y^\theta)$, we expand the brane action (\ref{braneact}) as
\begin{eqnarray}
S_{4D}&\supset&\int d^4x\sqrt{-g}\left(-f^4+\frac{m^2_X}{2\kappa^2}X_\mu X^\mu+\frac{\lambda}{\kappa^2}X^\mu X^\nu T^{\text{SM}}_{\mu\nu}\right.\nonumber\\
&&\left.+\mathcal{L}_{\text{SM}}+\cdots\right),\label{effbract}
\end{eqnarray}
where $T^{\text{SM}}_{\mu\nu}=2\delta\mathcal{L}_{\text{SM}}/\delta g^{\mu\nu}-g_{\mu\nu}\mathcal{L}_{\text{SM}}$ is the energy-momentum tensor of the SM.
With $X_\mu+\partial_\mu Y^\theta\rightarrow X_\mu$, the branon has been absorbed as the longitudinal mode of the KK gauge boson corresponding to the spontaneous breaking of the local isometry of $S^1$. As a result, the KK gauge boson acquires a mass $m_X=\sqrt{2}f^2/M_{\text{P}}$. In analogy with the condensate of the Higgs field in the Higgs mechanism, the formation of the $3$-brane whose tension determines the local density of the condensate would define a preferred position on the internal space $S^1$ and thus the KK gauge boson coupling directly to the excitations of the $3$-brane along $S^1$ would eat the branon to become massive. This process would correspond to a phase transition below a certain critical temperature where the $3$-brane is created. 

It is interesting that as seen in the effective actions (\ref{4Dgravact}) and (\ref{effbract}) the KK gauge boson is odd under the $Z_2$ symmetry which guarantees its stability. In addition, the KK gauge boson couples very weakly to the SM particles with the corresponding coupling given by $\lambda=1/M^2_{\text{P}}$ that is controlled by the 4D Planck scales like the usual gravitation because of the fact that the KK gauge boson has an origin of geometry associated with new gravitational degrees of freedom. Therefore, the KK gauge boson behaves actually as the DM, which suggests a geometric unification of gravity and the DM.

\section{Production during the inflation}
Because the coupling of the KK gauge boson to the SM particles is suppressed by the observed Planck scale, it must not have been produced in the thermal equilibrium. In this sense, the KK gauge boson DM would be the non-thermal relics and could be purely gravitationally created during the inflation due to the non-adiabatic expansion of the universe acting on the vacuum quantum fluctuations \cite{Ford1987,Lyth1998}. This production mechanism is particularly simple and interesting because it only bases on the Stueckelberg mass of the vector DM and the inflation which is an essential ingredient for the cosmology \cite{Graham2016}.

In order to compute the pure gravitational production of the KK gauge boson DM due to the quantum fluctuations during the inflation, first let us write the expansion law along the $3$-brane as \cite{Binetruy2000,Deffayet2000}
\begin{eqnarray}
H^2+\frac{k^2}{a^2}=\left(\frac{\rho_{\text{m}}+M^2_{\text{P}}m^2_X/2}{6M^3_5}\right)^2+\frac{\Lambda_5}{6},\label{NSHubb}
\end{eqnarray}
where the relation $f^4=M^2_{\text{P}}m^2_X/2$ has been used and $\rho_{\text{m}}$ is the matter-energy density on the $3$-brane. One can see that the standard expansion law is recovered in the low energy limit with $\rho_{\text{m}}\ll M^2_{\text{P}}m^2_X/2$ and the contribution of the brane tension compensated by the bulk cosmological constant by identifying $\Lambda_5=-6M^6_5/M^4_{\text{P}}$ leading to the mass of the KK gauge boson DM determined in terms of the compact radius as $m_X=\sqrt{3}/(\pi R_0)$.

Calculating the relic abundance of the KK gauge boson particles following Ref. \cite{Graham2016} with the non-standard expansion law (\ref{NSHubb}) and using the tensor power spectrum amplitude given in the brane-world scenario as $A_t\simeq12x_0/\left[H_i/(2\pi M_{\text{P}})\right]^2$ where $x_0\equiv2\sqrt{3H^2_i M^2_{\text{P}}/(2f^4)}$ and $H_i$ is the Hubble parameter during the inflation \cite{Langlois2000}, we find
\begin{eqnarray}
\frac{\Omega_X}{\Omega_{\text{CDM}}}&\simeq&1.33\times\left(\frac{rA_sM^2_{\text{P}}}{10^{26}\text{GeV}^2}\right)^{2/3}\left(\frac{m_X}{10^3 \text{GeV}}\right)^{7/6},\label{KKGBDM-rab}
\end{eqnarray}
where $\Omega_{\text{CDM}}h^2\simeq0.12$ is the observed DM relic density \cite{Planck2020} and $A_s\approx2.1\times10^{-9}$ is the scalar power spectrum amplitude \cite{Planck2020b}. In the present model, the ratio $r$ depends on the vector DM mass $m_{\text{DM}}$ as $r\propto m^{-7/4}_{\text{DM}}$, whereas it is $r\propto m^{-1/2}_{\text{DM}}$ for the dark vector DM models (with the same gravitational production mechanism) investigated previously \cite{Graham2016,Ema2019,Ahmed2020}. Because of this difference, the KK gauge boson DM model is distinguishable from the models of the dark vector DM in the literature and it also indicates a new region of the parameter space to search for the dark vector DM. For instance, with $r$ to order $0.01$ the models of the vector DM in the literature and the present model predict the vector DM mass of the order of $10^{-5}$ eV and $100$ GeV, respectively. As seen later, this new parameter region turns out to be consistent with quantum gravity.

\section{The constraint of Sublattice WGC}
We consider the constraint of Sublattice WGC on the scenario of the dark vector DM whose relic abundance is produced by the quantum fluctuations during the inflation. First, applying the constraint of this conjecture for the previously studied models of the relevant dark vector DM \cite{Graham2016,Ema2019,Ahmed2020} gives \cite{Reece2019}
\begin{eqnarray}
H_i\lesssim\Lambda_{\text{UV}}\lesssim g^{1/3}M_{\text{P}}\lesssim\left(\frac{4\pi m_A}{H_i}\right)^{1/3}M_{\text{P}},
\end{eqnarray}
where $g$ is the gauge coupling associated with the dark vector and $m_A$ is its mass. From the relic abundance of the dark vector DM found in \cite{Graham2016} and $r=2H^2_i/(\pi^2 A_sM^2_{\text{P}})$ associated with the standard expansion law, we find an upper bound on the tensor-to-scalar ratio $r$ as follows 
\begin{eqnarray}
r\lesssim5.35\times10^{-6}.
\end{eqnarray}
If the Sublattice WGC is true, this upper bound implies that the models of the dark vector DM in the literature which is consistently coupled to quantum gravity and is compatible with the DM observations must predict a tiny tensor-to-scalar ratio that is beyond the reach of the near future experiments. In this way, quantum gravity would wipe out an experimentally interesting region of the tensor-to-scalar ratio $r$, i.e $r\in\sim[10^{-4},10^{-2}]$ that is reachable at the current and future experiments, which is inconsistent with quantum gravity or belongs to the swampland.

However, it is interesting that the KK gauge boson DM model is safe from  the constraint of the Sublattice WGC. Indeed, from $H^3_i=\pi^2M^2_{\text{P}}m_XA_sr/(6\sqrt{3})$ with $m_X=\sqrt{3}/(\pi R_0)$ found above and Eq. (\ref{KKcoup}), we can show that the constraint $H_i\lesssim\Lambda_{\text{UV}}\lesssim \kappa^{1/3}M_{\text{P}}$ corresponds to $A_sr\lesssim6\sqrt{2}/\pi$ which is always satisfied with the current observations of $A_s$ and $r$. This means that, unlike the models of the dark vector DM in the literature, the value region of the tensor-to-scalar ratio which is experimentally accessible in the present or near future is not wiped out by the constraint of Sublattice WGC or belongs to the landscape of quantum gravity. In this way, the KK gauge boson DM model rescues the simple and intriguing DM production mechanism by only relying on the longitudinal mode of the vector DM and the inflation from the low-energy constraint of quantum gravity which excludes the experimentally interesting region of the relevant parameter space.

\section{Probing KK gauge boson DM in CMB and PGWs}

In the scenario of the dark vector DM whose relic abundance is generated purely gravitationally, the DM is generically coupled very weakly to the SM particles. This means that it is hard to probe its signatures at the colliders like LHC or in other words it is hard to confirm observationally this scenario. Hence, it is usually assumed a tiny kinetic mixing between the dark vector DM and the photon of the SM, leading to a non-gravitational interaction, for its detection in the experiments \cite{Graham2016}. Such a model of the dark vector DM has been searched for the predicted mass range of $\sim10^{-5}-10^{-1}$ eV in the laboratory experiments using the resonant LC oscillators \cite{Chaudhuri2015,Arias2015}. However, in our present model, the production of the KK gauge boson DM during the inflation happens at the very high energy scale which is much larger than $(M_{\text{P}}m_X)^{1/2}$ and thus the inflation is very sensitive to the presence of the KK gauge boson DM as seen from Eq. (\ref{NSHubb}). In this way, the KK gauge boson DM would leave imprints in the CMB as well as the PGW spectrum produced during the inflation. The accurate measurements of the CMB and PGWs would provide a means of probing the signatures of the KK gauge boson DM as well as another potential observation window to detect the scenario of the dark vector DM (produced purely gravitationally) in the mass range of the order of hundreds of GeV.

It is clear that the measurable signatures of the KK gauge boson DM depend on the inflationary model. In the present work, we consider the $\alpha$-attractor inflationary model, which is motivated by supergravity/string theory \cite{Kallosh2013} and favors the CMB data very well \cite{Planck2020b}, on the $3$-brane with the corresponding potential $V(\phi)$ given by
\begin{eqnarray}
V(\phi)=\Lambda^4\left(1-e^{-\sqrt{\frac{2}{3\alpha}}\frac{\phi}{M_{\text{P}}}}\right)^2,\label{infpot}
\end{eqnarray}
where $\Lambda$ and $\alpha$ are the parameters. Because the relevant expansion law in the presence of the KK gauge boson DM is $H\simeq V(\phi)/(3M_{\text{P}}m_X)$, its presence would cause an important change in the energy density at the end of the inflation and the $e$-folding number of the expansion during the reheating denoted by $N_{\text{re}}$. As a result, this affects the conversion of the inflaton energy into the radiation energy at the end of the reheating characterized by the reheating temperature $T_{\text{rh}}$. More specifically, we find the dependence of the reheating temperature $T_{\text{rh}}$ on the principal inflationary quantities observed in the CMB, namely, the scalar spectral index $n_s$, the tensor-to-scalar ratio $r$, the scalar power spectrum amplitude $A_s$ as follows
\begin{eqnarray}
\frac{T_{\text{rh}}}{\text{GeV}}&=&\left(\frac{45}{\pi^2g_*}\right)^{\frac{1}{4}}\frac{1.68\times10^{11}}{(A_sr)^{3/28}}\left[\sqrt{\beta(\beta+2)}-\alpha^2\right]^{\frac{1}{2}}\nonumber\\
&&\times \exp\left[-\frac{3}{4}(1+\omega_{\text{re}})N_{\text{re}}\right],\label{r-Trh-rel}
\end{eqnarray}
where we have used the correct DM abundance, $g_*$ is the effective number of relativistic degrees of freedom taken to be $\simeq106.75$ in this work, $\beta=[8(1-n_s)-3]/(4\sqrt{r})$, $\omega_{\text{re}}$ is the equation-of-state parameter which is zero for the inflationary potential (\ref{infpot}), and $N_{\text{re}}$ is given by
\begin{eqnarray}
N_{\text{re}}&=&\frac{4}{1-3\omega_{\text{re}}}\left[67.21-N_k-\ln\frac{k}{a_0H_0}-\frac{1}{12}\ln g_*\right.\nonumber\\
&&\left.+\frac{1}{4}\ln\frac{\pi A_sr}{27[\sqrt{\beta(\beta+2)}-\beta^2]^2}\right],
\end{eqnarray}
where $N_k=-(1/M^2_{\text{P}})\int^{\phi_f}_{\phi_i}V^2/(M^2_{\text{P}}m^2_XV')d\phi$, $k=0.002$ Mpc$^{-1}$ is the pivot scale, and $a_0$($H_0$) is the present value of the scale factor(Hubble parameter).

\begin{figure}[htp]
 \centering
\begin{tabular}{cc}
\includegraphics[width=0.4 \textwidth]{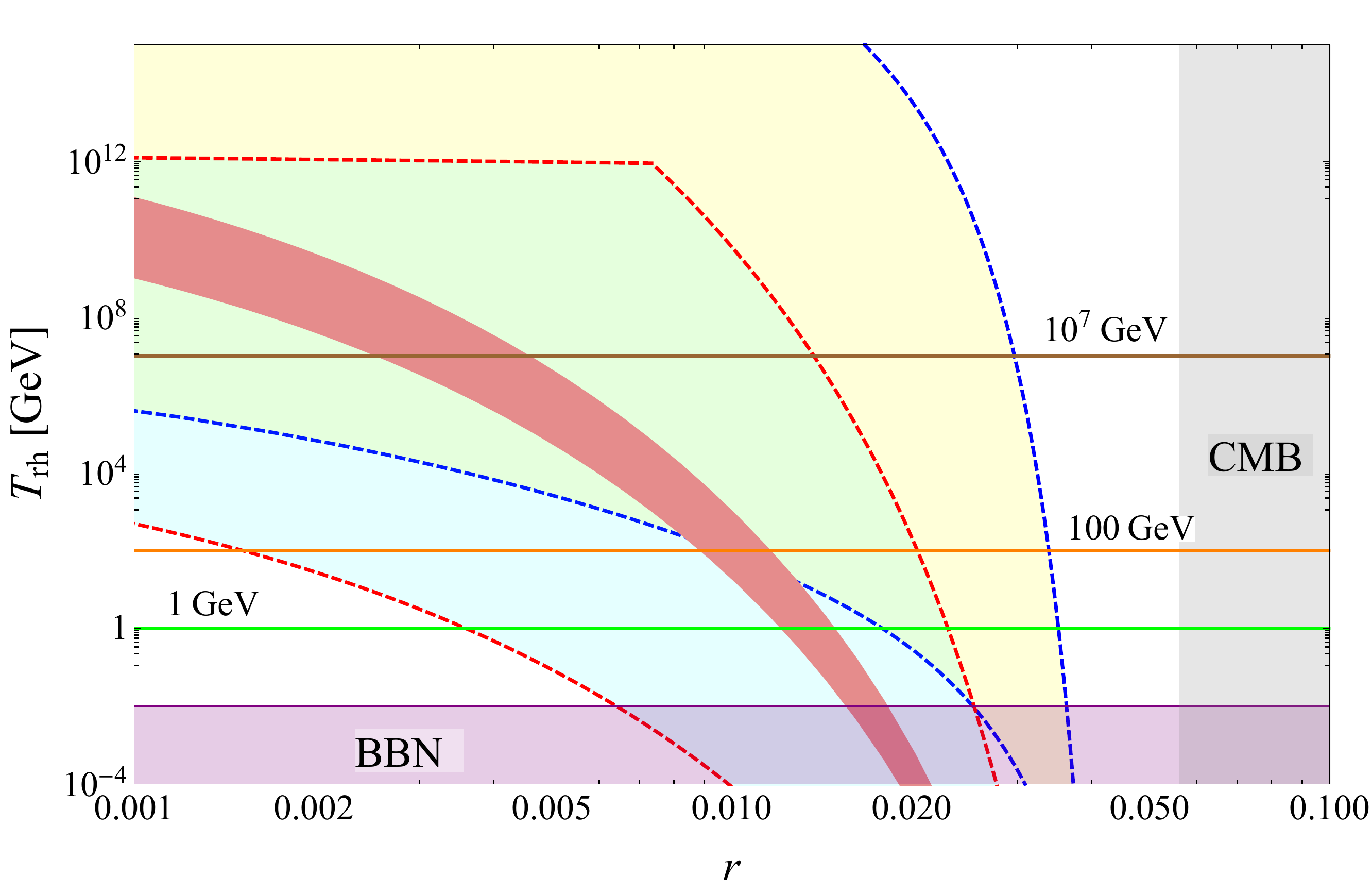}
\end{tabular}
 \caption{The prediction of the reheating temperature in terms of the tensor-to-scalar ratio in the KK gauge boson DM model, presented by a light cyan region bounded by the red dashed curves. The gray and light purple regions are excluded by the CMB and BBN constraints, respectively. The horizontal brown, yellow, and green lines refer to the upper bounds in the cosmological models based on supergravity \cite{Cyburt2003,Kawasaki2006,Felder2000}. We also include the prediction of the standard $\alpha$-attractor inflationary model for comparison, presented by the light yellow region bounded by the blue dashed curves.}\label{r-Tre}
\end{figure}

Fig. \ref{r-Tre} shows the prediction of our model for the reheating temperature in terms of the tensor-to-scalar ratio using the observation of the scalar spectral index $n_s=0.965\pm0.004$ ($68\%$ CL) \cite{Planck2020}, which is presented by the light cyan region bounded by the red dashed curves. The allowed interval of the reheating temperature is about $10$ MeV $-$ $10^{12}$ GeV. 
The reheating temperature above about $10^{12}$ GeV is unphysical because it leads to $N_{\text{re}}<0$. Because of the uncertainty on $n_s$, the prediction of the dependence of the reheating temperature $T_{\text{rh}}$ on the tensor-to-scalar ratio $r$ is currently less precise. However, the future CMB experiments \cite{Amendola2013,Andre2013} which will be able to decrease the error of the measurement of $n_s$ would allow a more precise dependence of $T_{\text{rh}}$ on $r$. For instance, we present the pink region assuming an uncertainty of $\sim5\times10^{-4}$ for $n_s$ and its unchanged central value. For comparison, we show the prediction of the standard $\alpha$-attractor inflationary model as the light yellow region bounded by the blue dashed curves. 

\begin{figure}[htp]
 \centering
\begin{tabular}{cc}
\includegraphics[width=0.4 \textwidth]{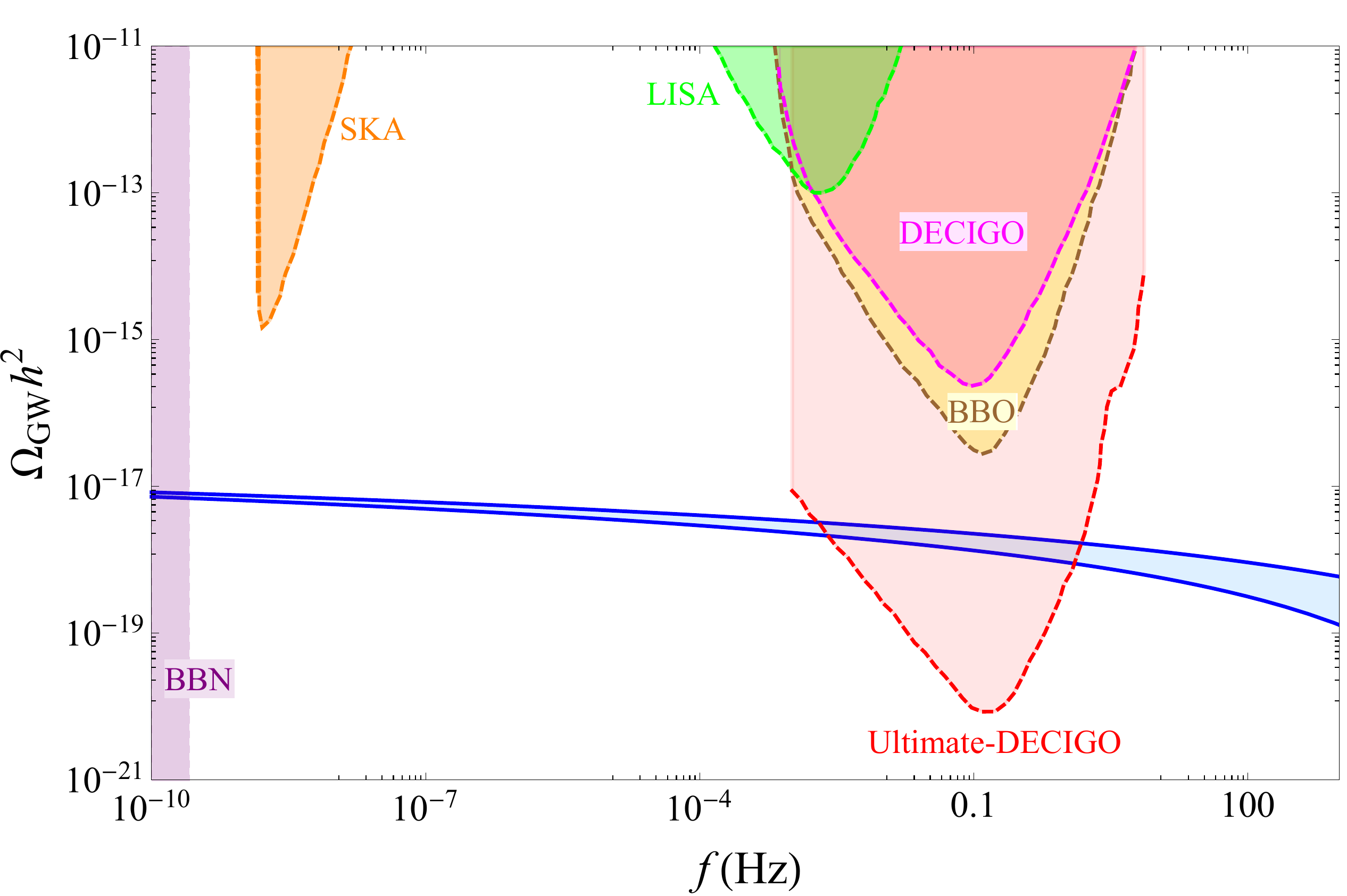}
\end{tabular}
\caption{The PGW signals of the KK gauge boson DM produced during the inflation detected at the reheating frequency, presented by the light blue region, within the reach of the future GW observations.}\label{f-OmWh}
\end{figure}

We observe from Fig. \ref{r-Tre} that the observed relic abundance of the KK gauge boson DM connected to the $\alpha$-attractor inflation leads to a low reheating temperature, which is a distinguishing feature of our model compared to the ordinary inflationary models: 

First, this makes it very predictive and testable from the accurate measurements of the spectral index $n_s$, the tensor-to-scalar ratio $r$, and the reheating temperature. The value of $r$ is measured from the CMB B-mode polarization which has been detected yet and hence $95\%$ CL upper limit on $r$ is imposed as $r<0.056$ \cite{Planck2020b}. Whereas, the reheating temperature can be determined from the spectrum of the inflationary GWs denoted by $\Omega_{\text{GW}}h^2$ corresponding to the detection frequency $f_{\text{rh}}$ as \cite{Kuroyanagi2011}
\begin{eqnarray}
f_{\text{rh}}\simeq0.26\text{Hz}\left(\frac{g_*}{106.75}\right)^{1/6}\frac{T_{\text{rh}}}{10^7\text{GeV}},
\end{eqnarray}
where the frequency dependence of $\Omega_{\text{GW}}h^2$ changes due to the occurrence of the reheating. From this relation together with Eq. (\ref{r-Trh-rel}) and the dependence of $\Omega_{\text{GW}}h^2$ on $r$, $T_{\text{rh}}$, and the frequency of the GWs, we depict the predicted values of $\Omega_{\text{GW}}h^2$ in our model in terms of the frequency where the reheating is detected, as the light blue region (corresponding to the light pink region in Fig. \ref{r-Tre}) in Fig. \ref{f-OmWh}. Additionally, we have included the proposed future GW observations like SKA \cite{SKA}, LISA \cite{LISA}, BBO \cite{BBO}, DECIGO \cite{DEC}, and Ultimate-DECIGO whose sensitivity is only limited by the quantum noises \cite{UDEC}. We see that the KK gauge boson DM can be probed by Ultimate-DECIGO in the range of the reheating frequency $\sim[2.2\times10^{-3},1.48]$ Hz corresponding to the range of the reheating temperature $\sim[8.5\times10^4,5.7\times10^7]$ GeV or the range of the tensor-to-scalar ratio $r\sim[3\times10^{-3},6\times10^{-3}]$ which is experimentally accessible in the near future.

Second, the region of the low reheating temperature here would be of particular interest in constructing the cosmological models based on supergravity which requires $T_{\text{rh}}\lesssim10^7$ GeV in order to avoid the overproduction of gravitinos \cite{Cyburt2003} which has been a major obstacle for building such cosmological  models. In particular, it is indicated in the situation of very light gravitinos, i.e. $10^{-3}$ MeV $\lesssim m_{3/2}\lesssim$ $10^{-1}$ MeV, the reheating temperature has to be lower than $100$ GeV \cite{Kawasaki2006}. In addition, there is another challenge in the cosmological models based on supergravity is the increase of entropy arising from the production of moduli fields, which can be avoided if $T_{\text{rh}}<1$ GeV \cite{Felder2000}. With the current CMB data, our model consists of a wide enough range of parameters for incorporating supergravity, as seen in Fig. \ref{r-Tre}.

\section{Conclusion}

The compactification of superstring/M theory which is regarded as a consistent theory of quantum gravity leads naturally to the effective low-energy theories with gravity propagating in the extra dimensions and the SM fields confined to a $3$-brane, as well-known as the brane-world scenario. It was argued that the fluctuations of the $3$-brane along the extra dimensions are a natural DM candidate, so-called the branon DM studied extensively in the literature. However, in this work we show that the branons belong to the low-energy particle spectrum only in the decoupling limit of the KK gauge fields, which would result in the violation of the swampland conjectures. In this sense, the branon DM cannot come from quantum gravity or the string compactification. On the other hand, in order for the brane-world scenario to be UV completed in quantum gravity, the presence of the KK gauge bosons is avoidable and as a result, they would eat the branons to acquire a mass. 

We indicate a geometric unification of gravity and the DM where the role of the DM is played by the KK gauge bosons which are stable by the parity on the brane and couple very weakly to the SM particles due to their geometric nature and gravitational origin. The KK gauge boson DM is sensitive to the very high energy scales like the inflation scale and thus can be probed by the observations of the very early universe. We provide its predictions in the future observations of CMB and PGW spectrum produced during the inflation, which provides a complementary window to detect the DM coupling very weakly to the SM in a new parameter range of the DM mass in addition to the laboratory experiments.

\end{document}